\documentclass[aps,prl,preprint,superscriptaddress,longbibliography]{revtex4-2}
\usepackage{amsmath,latexsym}
\usepackage{xcolor}
\usepackage{graphicx}
\usepackage{caption}
\usepackage{color}
\usepackage{dcolumn}
\usepackage{bm}
\usepackage{CJK}
\usepackage{braket}
\usepackage[font=small,labelfont=bf]{caption}
\usepackage{amssymb}
\usepackage[colorlinks,linkcolor=blue,anchorcolor=blue,citecolor=blue]{hyperref}
\usepackage{epstopdf}
\usepackage{subcaption}
\usepackage[utf8]{inputenc} 
\usepackage[T1]{fontenc}    
\usepackage{url}            
\usepackage{booktabs}       
\usepackage{amsfonts}       
\usepackage{nicefrac}       
\usepackage{microtype}      
\usepackage{lipsum}
\usepackage{multirow}
\usepackage{fancyhdr}       
\graphicspath{{media/}}     

\begin{document}

\title{Local environment-based machine learning for molecular adsorption energy prediction}

\author{Yifan Li}
\affiliation{Department of Mechanical Engineering, National University of Singapore, Singapore 117542, Singapore}

\author{Yihan Wu}
\affiliation{Department of Mechanical Engineering, National University of Singapore, Singapore 117542, Singapore}

\author{Yuhang Han}
\affiliation{Department of Mechanical Engineering, National University of Singapore, Singapore 117542, Singapore}

\author{Qiujie Lyu}
\affiliation{Department of Mechanical Engineering, National University of Singapore, Singapore 117542, Singapore}

\author{Hao Wu}
\affiliation{Department of Mechanical Engineering, National University of Singapore, Singapore 117542, Singapore}

\author{Xiuying Zhang}
\affiliation{Department of Mechanical Engineering, National University of Singapore, Singapore 117542, Singapore}

\author{Lei Shen}
\email{shenlei@nus.edu.sg}
\affiliation{Department of Mechanical Engineering, National University of Singapore, Singapore 117542, Singapore}

\date{\today}

\begin{abstract}
Most machine learning (ML) models in Materials Science are developed by global geometric features, often falling short in describing localized characteristics, like molecular adsorption on materials. In this study, we introduce a local environment framework that extracts local features from crystal structures to portray the environment surrounding specific adsorption sites. Upon OC20 database (~20,000 3D entries), we apply our local environment framework on several ML models, such as random forest, convolutional neural network, and graph neural network. It is found that our framework achieves remarkable prediction accuracy in predicting molecular adsorption energy, significantly outperforming other examined global-environment-based models. Moreover, the employment of this framework reduces data requirements and augments computational speed, specifically for deep learning algorithms. Finally, we directly apply our Local Environment ResNet (LERN) on a small 2DMatPedia database (~2,000 2D entries), which also achieves highly accurate prediction, demonstrating the model transferability and remarkable data
efficiency. Overall, the prediction accuracy, data-utilization efficiency, and transferability of our local-environment-based ML framework hold a promising high applicability across a broad molecular adsorption field, such as catalysis and sensor technologies.
\end{abstract}

\maketitle


\section{Introduction}
Machine learning (ML) has found widespread applications in the field of Materials Science and Engineering\cite{sha2020artificial,choudhary2022recent}. ML techniques, such as SchNet\cite{schutt2017schnet}, CGCNN\cite{xie2018crystal}, and Alignn\cite{choudhary2021atomistic}, have been employed to establish relationships between atomic structures and their properties. These methods have been utilized to predict more than fifty different characteristics of crystals and molecular materials, including formation energy and electronic band gaps. Moreover, atomic neural networks have been instrumental in developing interatomic potentials\cite{behler2007generalized} or Hamiltonian\cite{zhong2023transferable} instead of direct predictions of physical properties. Additionally, deep learning techniques have been utilized in various applications to identify chemically feasible spaces. For instance, Bayesian optimization methods, in conjunction with MEGNet, have been employed as energy evaluators for direct structure relaxation\cite{chen2019graph}. To further enhance the performance, BOWSR incorporates band symmetry relaxation alongside Bayesian optimization\cite{zuo2021accelerating}.

Accurate characterization of localized physicochemical properties is of paramount importance in numerous scientific and engineering disciplines. Ranging from electrochemical catalysis, sensors, carbon capture, energy storage and conversion, to drug delivery, exploring and exploiting the intricacies of local environments are at the heart of many frontier investigations. For instance, adsorption is a pervasive surface phenomenon in areas like electrocatalysis, with its understanding rooted in foundational theories such as bonding and adsorption thermodynamics\cite{conway2002interfacial}. Crucially, the influence of neighboring atoms on adsorption sites must be fully accounted for, as factors including atomic electronic structures, spatial constraints, surface stoichiometry, and surface defects can all impinge on the behavior of adsorbates on surfaces\cite{loffreda2006theoretical}. 

For such localized concerns, earlier studies proposed numerous feature engineering descriptors to enhance the prediction of ML models for adsorption energy, encompassing atomic number, ionization energy, electronegativity, ionic radius, and inter-atomic interactions\cite{isayev2015materials, ward2016general, schutt2014represent}. These descriptors are equally pertinent to the field of electrocatalysis. Yet, adequately representing metal compound adsorption sites remains a major challenge. Tran and Ulissi introduced the use of geometric fingerprints to depict the local region around each atom, which were restricted to rudimentary elemental properties\cite{tran2018active}. Recently, a variety of end-to-end graph neural network models have been developed in the OCP challenge, encompassing equivariant e3nn networks, SCN, and eSCN, among others\cite{thomas2018tensor, batzner20223, brandstetter2021geometric, batatia2022mace, musaelian2023learning, liao2022equiformer, zitnick2022spherical, passaro2023reducing,lan2023adsorbml,wang2022heterogeneous,zhong2023transferable}. While these models showcase stellar performance on the data-rich OC20 database\cite{chanussot2021open}, their computational complexity often leads to overfitting on smaller datasets. Specifically, for certain complex adsorption systems like large organic molecules and certain transition metal oxides, accurate density functional theory (DFT) computations are challenging, resulting in a paucity of reliable data\cite{liu2014modeling, shee2019achieving}. The Adsorbate Chemical Environment-based Graph Convolution Neural Network (ACE-GCN) endeavors to capture the local information of adsorbates using molecular graphs, introducing incremental improvements to address this issue. Additionally, employing ML approaches boasts advantages such as heightened interpretability, reduced data dependency, flexibility in designing and selecting features tailored to specific problems, and capabilities in prediction interpretation and error analysis \cite{yang2022applications, george2021chemist, zebari2020comprehensive, otchere2021application,huang2023exploring,li2023graph}. Very recently, feature engineering based on central environments has demonstrated efficacy in describing local environments\cite{li2023center}, but facing generality and transferability issues and proving not universally applicable to surfaces\cite{chen2022combined}.

In light of the above backdrop, this work embarks on exploring a novel local environment-based approach for adsorption-energy prediction. We not only explore deep into the detailed description of the local chemical environment, but also strive to overcome the limitations of existing methodologies. The method we introduce marries the robust performance of deep learning with the flexibility and interpretability of machine learning, aiming to accelerate the training process and enhance predictive accuracy across multiple domains. Ultimately, this work paves the way for new possibilities in understanding and manipulating the complexity of local environments.

\begin{figure}%
\centering
\includegraphics[width=1\textwidth]{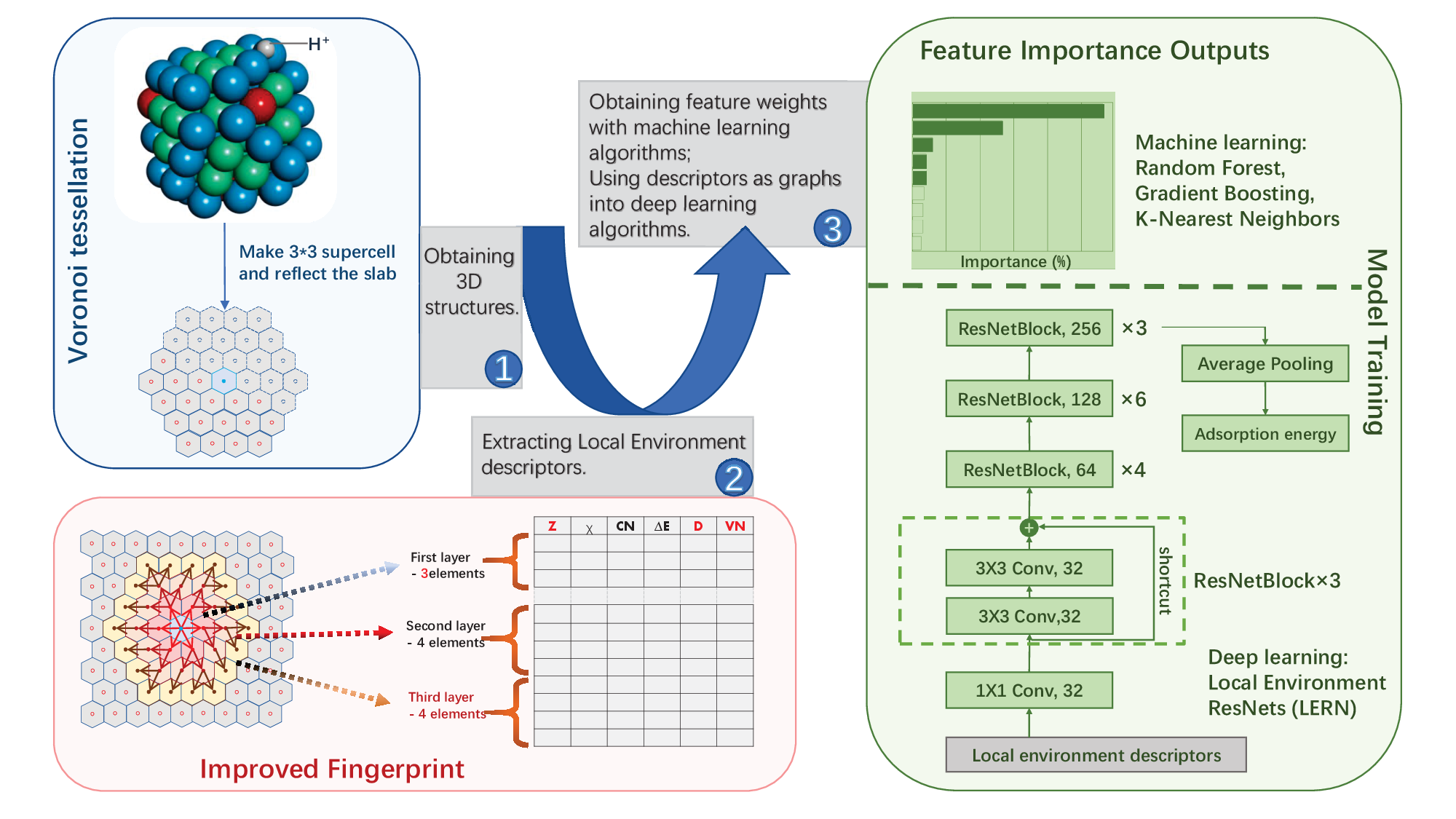}
\caption{{The general framework of the local environment model. The method provides a descriptor-driven solution for adsorption calculations, especially for large surfaces, without the need for extensive structure-driven training resources. Given the generalizability of the descriptor extraction method, trained models are expected to be able to predict the sorption energy of other sorbents on a variety of complex surfaces. The method is also highly interpretable and retains the advantages of traditional machine learning algorithms.}}
\label{fig:ALL}
\end{figure}

\section{Methods}

The concept of local environment, for machine learning-based prediction of adsorption, is primarily rooted in the feature extraction of the surrounding environment of the adsorbate. We first locate the position of the adsorbate within the three-dimensional structure, and then sequentially extract the atoms neighboring the adsorbate at each layer (Fig. 1 Step 1). Through feature processing of the surrounding atoms at each layer, we obtain descriptors that describe the local environment (Fig. 1 Step 2). This approach allows us to capture the essential information about the adsorbate's immediate surroundings and characterize its local environment effectively. Figure 1 (Step 1) shows an optimized Voronoi Tessellation (VT) technique, which is used in this work, to the surface adsorption system (see more details of the optimized VT in SI). The optimized VT method offers several advantages, including freedom from parameter tuning, symmetry and composition independence, transferability, and reproducibility. 

After transforming the original material structure into a Voronoi diagram, we extract improved local environment fingerprint information. The new features, different from conventional fingerprint methods, are highlighted in red in Fig. 1 (Step 2). Our final results prove that only six descriptors make the conventional ML models reach a significantly high accuracy (Table 1). The feature details and comparison with the conventional VT model can be found in Supporting Information. 

Besides for conventional ML algorithms, our local environment features can also be transformed into a 6*11 matrix, which meets all the necessary criteria for neural network utilization. By doing so, we can leverage the advantages of deep learning in the context of adsorption energy calculations. In this work, we use the ResNet-34 model as our backbone neural network (Fig. 1 Step 3). ResNet-34 consists of 33 convolutional layers, providing depth and capability to the model, enabling it to be applied to larger databases, thereby enhancing the model's accuracy and robustness. Overall, by constructing a local environment convolutional network, we integrate the high-importance features of the three nearest neighbor atoms surrounding the adsorbate and learn the relationships between each layer's elements. This local environment framework is general for both conventional ML algorithms and deep learning neural networks, for both large datasets and small datasets, and for both 3D and 2D materials. Performance metrics for all methods used in this article are Median Absolute Error (MDAE), Mean Absolute Error (MAE), Root Mean Squared Error (RMSE), Mean Absolute Relative Percent Difference (MARPD). Our local environment-based models show significant accuracy and efficiency compared with conventional global environment ML models.

\section{Results}

\subsection{Datasets}

This work utilize two computational material databases for training, comparison, validation and generalization, a big-data 3D material-based adsorption energy dataset (from OC20) and a small 2D material-based adsorption energy dataset (from 2DMatPedia).  The 3D material datasets comprise a remarkable 47,279 DFT-calculated adsorption energy values. These calculations were conducted using the Generalized Adsorption Simulator for Python\cite{tran2018dynamic}\cite{tran2018active}. The dataset includes 21,269 adsorption energies concerning hydrogen atoms, which are the central dataset of this study. Additionally, there are 26,010 adsorption energies pertaining to other atoms. This extensive dataset covers a wide range of 52 chemical elements and 1,952 bulk materials, thereby enhancing its relevance and applicability. Furthermore, it is enriched with 9,102 symmetrically distinct surfaces and 29,843 distinct coordination environments, all carefully characterized based on the surface and the adsorbate neighbors.The 2D material database contains a substantial collection of 2,472 DFT calculations for hydrogen-atom adsorption energy\cite{yang2020high}. These calculations are performed on surfaces obtained from our developed 2Dmatpedia database\cite{zhou20192dmatpedia}, which currently encompasses over 10,000 distinct 2D materials. The utilization of these comprehensive and diverse databases ensures that our study's findings are both robust and pertinent, paving the way for significant contributions to the field. The data distribution is largely normal and is therefore deemed suitable for machine learning methods. The training set comprises 80\% of the dataset, while the test and validation sets account for 10\% each.

\subsection{Local Environment-based Conventional ML}

Though the low accuracy compared to DL, conventional machine learning algorithms can output weights for descriptors and have high interpretability. Thus, they are often used to evaluate the accuracy of the improved descriptors\cite{cao2019convolutional}. Here, we first applied our local environment features on three widely-used ML algorithms, including in Gradient Boosting (GB)\cite{friedman2002stochastic}, K-Nearest Neighbors (KNN)\cite{cover1967nearest}, and Random Forest (RF) algorithms\cite{breiman2001random}, following data filtering. We employed the RF algorithm initially to determine the feature importance in the training process, refer to Appendix A.3 for specific results upon further analysis.

Table{ref{tab:ML} summarises the training results. The random forest model incorporating local environment descriptors produces the lowest MAE value of 0.13 eV. The RF algorithm outperforms the other two machine learning methods thanks to its ability to integrate decision trees and capture complex inter-atomic relationships. In addition, the same ML algorithms equipped with our local environment framework outperform those with the Voronoi Tessellation (VT) method due to its inability to incorporate information about adsorbed atoms into the model. As a result, the local environment descriptor methods showed reliable performance in predicting the adsorption energy processes with significantly lower error rates.

\begin{table}[h]\centering
\begin{center}
\begin{minipage}{\textwidth}
\caption{Performance comparison of machine learning models using different descriptors}
\label{tab:ML}
\begin{tabular*}{\textwidth}{@{\extracolsep{\fill}}lcccccc@{\extracolsep{\fill}}}
\toprule%
\begin{tabular}{@{}l@{}}
 Modeling\\ method \\
\end{tabular}
&
\begin{tabular}{@{}l@{}}
Learning\\{ method}\\
\end{tabular}
&MDAE&MAE&RMSE&MARPD\\
\midrule
\multirow{3}*{
\begin{tabular}{@{}l@{}}
 Voronoi\\Tessellations \\
\end{tabular}}
 & GB &0.20 & 0.28& 0.42 &108\% \\
 & KNN  &0.20 &  0.29 & 0.44 &105\% \\
 & RF  &0.19 &\textbf{0.27} & 0.41 &103\% \\
\midrule
\multirow{3}*{Improved Fingerprint}
 & GB  &0.16 & 0.20 & 0.27 &89\% \\
 & KNN &0.11 &  0.17 & 0.25 &72\% \\
 & RF  &0.08 &\textbf{0.13} & 0.22 &61\% \\
\bottomrule
\end{tabular*}
\end{minipage}
\end{center}
\end{table}

\subsection{Local Environment-based CNN}

\subsubsection{Molecular CGCNN}

Besides the accuracy, we next will discuss the data efficiency of our local environment framework. Deep learning algorithms, especially convolutional neural network (CNN), have shown their great advances and applications in materials science. One of the famous DL methods in materials science, Crystal Graph Convolutional Neural Network (CGCNN) deep learning framework, has demonstrated remarkable performance in several applications. In this study, we propose the use of our modified VT\cite{tanemura1983new} method to optimize the original structure of CGCNN into a Voronoi structure input, which we name the Molecular Crystal Graph Convolutional Neural Network (MolCGCNN) as shown in Fig.\ref{fig:molCGCNN}. This approach shows that our modified VT method gives a significant develop in the field of small data limit.

\begin{figure}
     \centering
     \begin{subfigure}[b]{0.8\textwidth}
         \centering
         \includegraphics[width=0.8\textwidth]{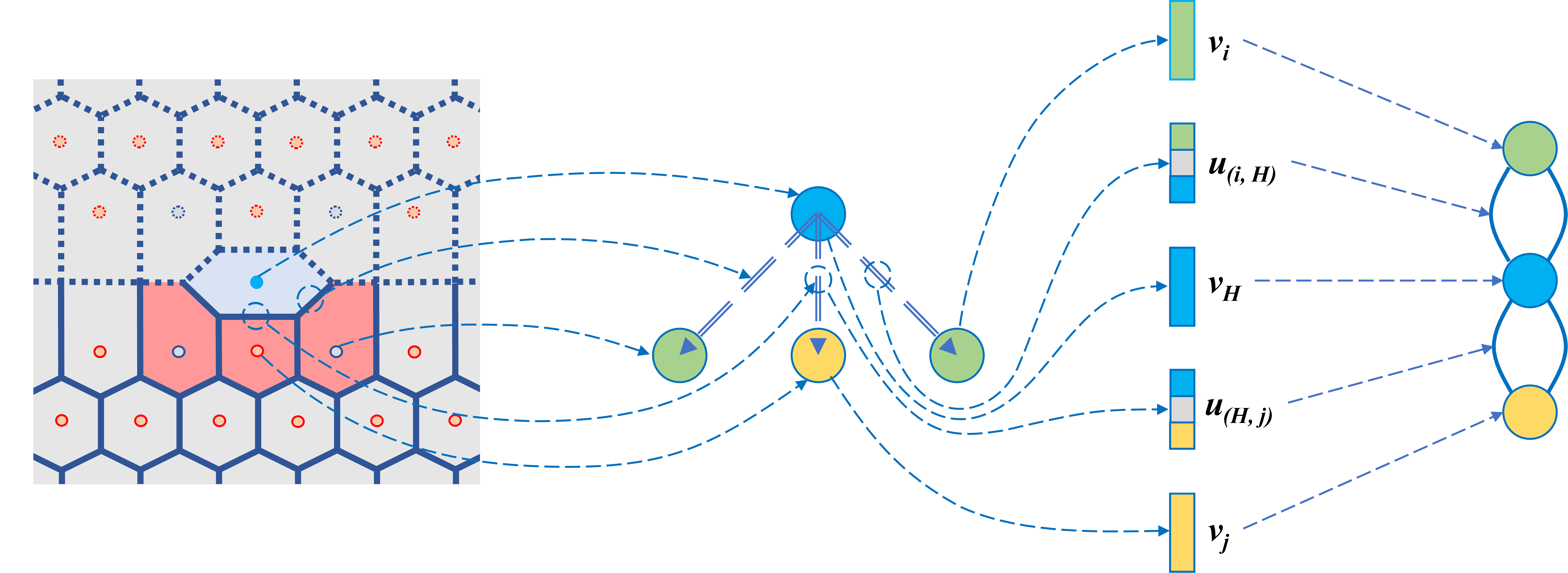}
         \caption{}
         \label{fig:molCGCNN}
     \end{subfigure}
     \hfill
     \begin{subfigure}[b]{0.9\textwidth}
         \centering
         \includegraphics[width=1\textwidth]{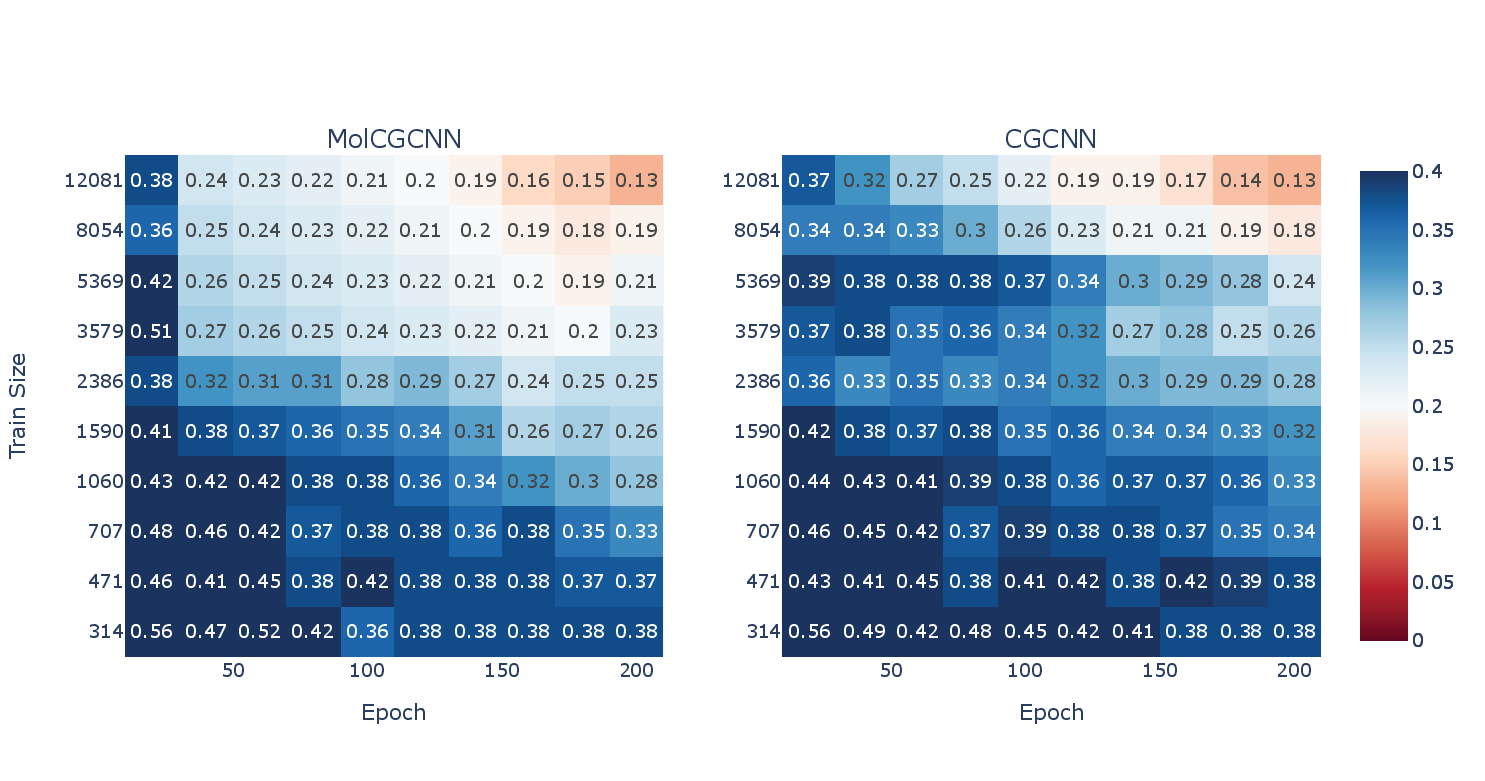}
         \caption{}
         \label{fig:molre}
     \end{subfigure}
     \hfill
     \caption{(a) Illustration of MolCGCNN crystal graph. (b)The heatmaps of two models of MAE(eV) corresponding to different training sizes and Epochs. }
\end{figure}

Our results, as depicted in Fig.\ref{fig:molre}, show that MolCGCNN achieves superior convergence performance compared to the original CGCNN. The proposed model efficiently captures the necessary information for calculating the adsorption energy with fewer iterations. We attribute this to the enhanced filtering of features by extracting local information, thus facilitating the identification of information localized to the adsorption site.  The faster training speed of MolCGCNN is particularly relevant for large datasets, given that deep learning algorithms typically require long training cycles due to the high number of hidden layers in neural networks.In Appendix A.2, we present a detailed account of our findings and provide supporting evidence for our claims.However, there are some inherent limitations of crystal graph networks in adsorption energy calculations, which we discuss in the Methods section, and therefore we consider the application of feature engineering in this area.

\subsubsection{Local Environment ResNet}

Finally, to achieve both accuracy and efficiency, we utilize a convolutional neural network with a 3$\times$3 convolutional kernel to further process this graph, mapping the matrix to predicted parameters, specifically adsorption energy. To enable the use of deeper networks, we introduce a residual network (ResNet)\cite{he2016deep} into our local environment model. The ResNet has excellent tunability and fast training speed, allowing for greater versatility in our model's application to a wider range of adsorbates and scenarios. Additionally, the ResNet's strong generalization performance facilitates easy portability of our model to other fields. In this model, we use mean absolute error as the loss function and Adam as the optimization algorithm to train the network.

The detailed structure is shown in Table.\ref{tab:str}.

\begin{table}[h]\centering
\begin{center}
\begin{minipage}{\textwidth}
\caption{The structure of Local Environment ResNet}
\label{tab:str}
\begin{tabular*}{\textwidth}{@{\extracolsep{\fill}}lcccccc@{\extracolsep{\fill}}}
\toprule%
Layer Name&Output Size
&34-Layer\\
\midrule
Conv1 & $6\times11$ &$1\times1,32$, stride 1 \\

\multirow{2}*{Conv2\_x} & \multirow{2}*{$6\times11$} &$ \left[ \begin{matrix}
3\times3,32\\
3\times3,32
\end{matrix}\right]\times3 $, stride 1 \\

\multirow{2}*{Conv3\_x} & \multirow{2}*{$3\times6$} &$ \left[ \begin{matrix}
3\times3,64\\
3\times3,64
\end{matrix}\right]\times4 $, stride 2 \\

\multirow{2}*{Conv4\_x} & \multirow{2}*{$2\times3$} &$ \left[ \begin{matrix}
3\times3,128\\
3\times3,128
\end{matrix}\right]\times6 $, stride 2 \\

\multirow{3}*{Conv5\_x} & $1\times2$&$ \left[ \begin{matrix}
3\times3,256\\
3\times3,256
\end{matrix}\right]\times3 $, stride 2  \\
&$1\times1$
&AdaptiveAvgPool \\
\bottomrule
\end{tabular*}
\end{minipage}
\end{center}
\end{table}

The prediction results of the LERN model are shown in Fig.\ref{fig:lernre}. The Residual plot is beneficial for analyzing the distribution of prediction errors. In the scatter plot on the left, the x-axis and y-axis represent the LERN training process and the residuals between predicted and actual values on both the training and test sets, respectively. The overall residual regression line is also calculated and displayed in the plot. The red and blue parts represent the training and test sets, respectively. It can be observed that the LERN prediction results are highly consistent with the actual values on both the training and test sets, as evidenced by the residuals' regression lines with slopes close to zero. On the right, the overall distribution of residuals on the training and test sets is displayed. The results show that the distribution of residuals on both the training and test sets is close to normal distribution, indicating that the model is well-suited for learning from this type of data. Moreover, the distributions of residuals on the training and test sets are very similar, indicating that the model does not suffer from obvious overfitting. Therefore, the LERN prediction results are highly consistent with the actual DFT calculation results within the allowable error range, suggesting that LERN can be used to replace DFT calculations.

\begin{figure}
\centering
\includegraphics[width=0.8\textwidth]{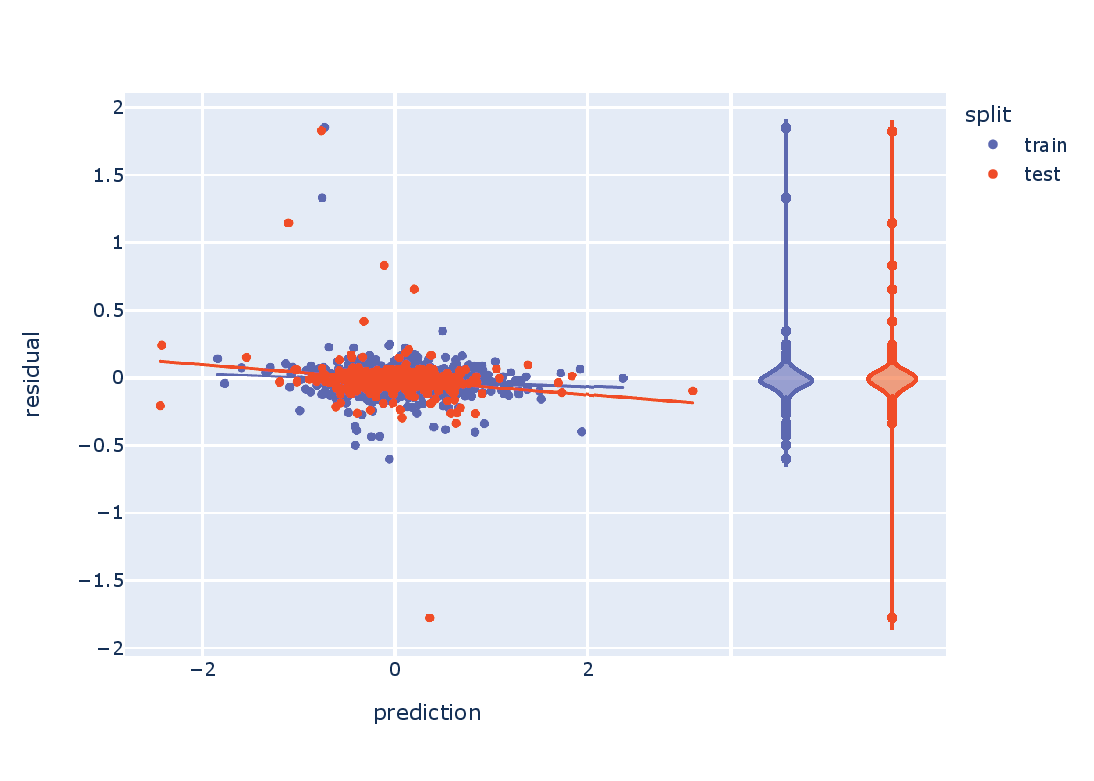}
\caption{Residual plot of LERN models. The left-hand side is a scatter plot, where the x-axis is the predicted value of the model on the training and test sets, the y-axis is the corresponding predicted value minus the true value, and the middle corresponds to its regression line respectively. On the right-hand side are the residual distributions for the training and test sets, respectively.}
\label{fig:lernre}
\end{figure}

\begin{figure}
\centering
\includegraphics[width=1.\textwidth]{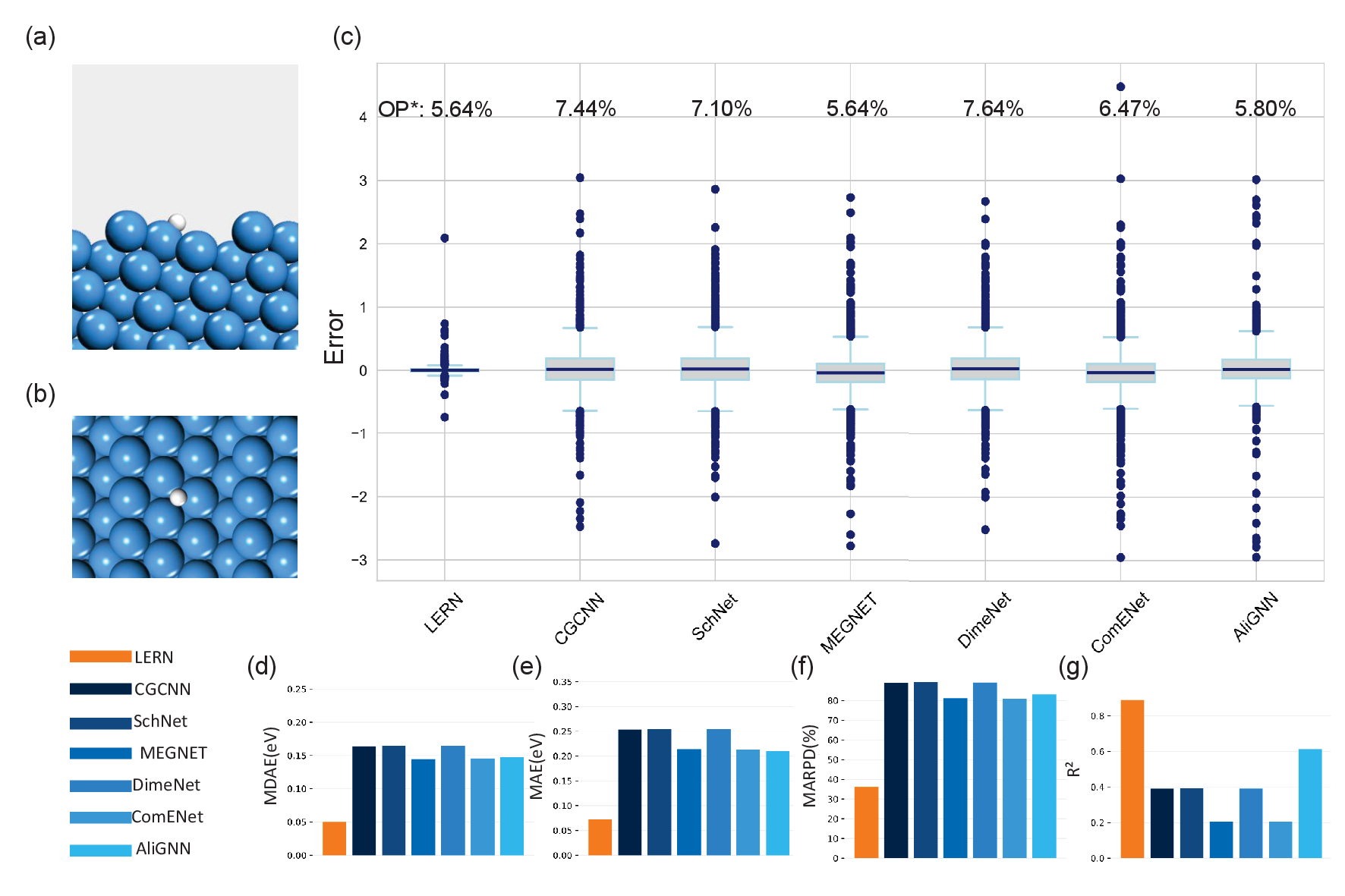}
\caption{(a)H adsorbed on 3D materials (b) and (c)Performance comparison of LERN with other representative models on 3D materials.(*op:Outlier percentage)}
\label{fig:box}
\end{figure}

To benchmark the performance of our proposed model in a materials database, we use other state-of-the-art neural networks. We compare our Local Environment ResNet (LERN) with the original Crystal Graph Convolutional Neural Network (CGCNN) and SchNet\cite{schutt2018schnet} and Atomistic Line Graph Neural Network(AliGNN)\cite{choudhary2021atomistic}, as well as our improved versions of the Molecular CGCNN. To elaborate further, the robustness of LERN in dealing with outliers is a highly desirable trait in machine learning models. Outliers are data points that deviate significantly from the normal distribution of the dataset, and they can occur due to various reasons such as measurement errors or anomalous samples. The presence of outliers can negatively impact the performance of a model, especially if it is not designed to handle them properly. In contrast, as shown in Fig.\ref{fig:box}, the orange points represent the distribution of outliers for each model. Of these, LERN has only 86 outliers, while the other models are all around 110. This suggests that LERN has demonstrated an impressive ability to handle outliers effectively, which is a crucial advantage in real-world applications where data quality is often suboptimal. By being insensitive to outliers, LERN can deliver reliable predictions even in the presence of noisy data. This is particularly relevant in the field of materials science, where experimental data can be scarce, noisy, or incomplete, making it challenging to develop accurate models. Moreover, the performance of LERN on limited training data samples is noteworthy. Discrete errors often arise from inadequate training data or inherent low similarity in the training set, and the error distribution of LERN is more concentrated, indicating higher prediction accuracy and consistency. The ability to learn from a small amount of data is an important aspect of machine learning, as it allows for the development of models that can be trained with fewer computational resources and time. This is particularly important in fields such as materials science, where experiments can be both expensive and time-consuming. The ability of LERN to effectively learn from limited data suggests that it has the potential to significantly accelerate the discovery and design of new materials.

 Figure \ref{fig:box} shows that LERN surpasses other models in predicting the adsorption energy of HER. During our training, we discovered that, akin to MolCGCNN, LERN's accuracy remains stable even with fewer iterations and smaller sample sizes, whereas other models show significant deterioration under low data conditions. This finding reiterates that local environment descriptors are more fitting for limited data scenarios, reflecting the current state of most catalytic databases. Such high data efficiency can be attributed to the model's ability to extract vital local distance information and atomic properties from the structure, embedding system knowledge and input-output correlations. LERN is able to rapidly concentrate on the features surrounding the adsorption site using the training dataset, whereas other models must laboriously learn all atomic correlations without the benefit of system knowledge.

\begin{figure}
\centering
\includegraphics[width=1.\textwidth]{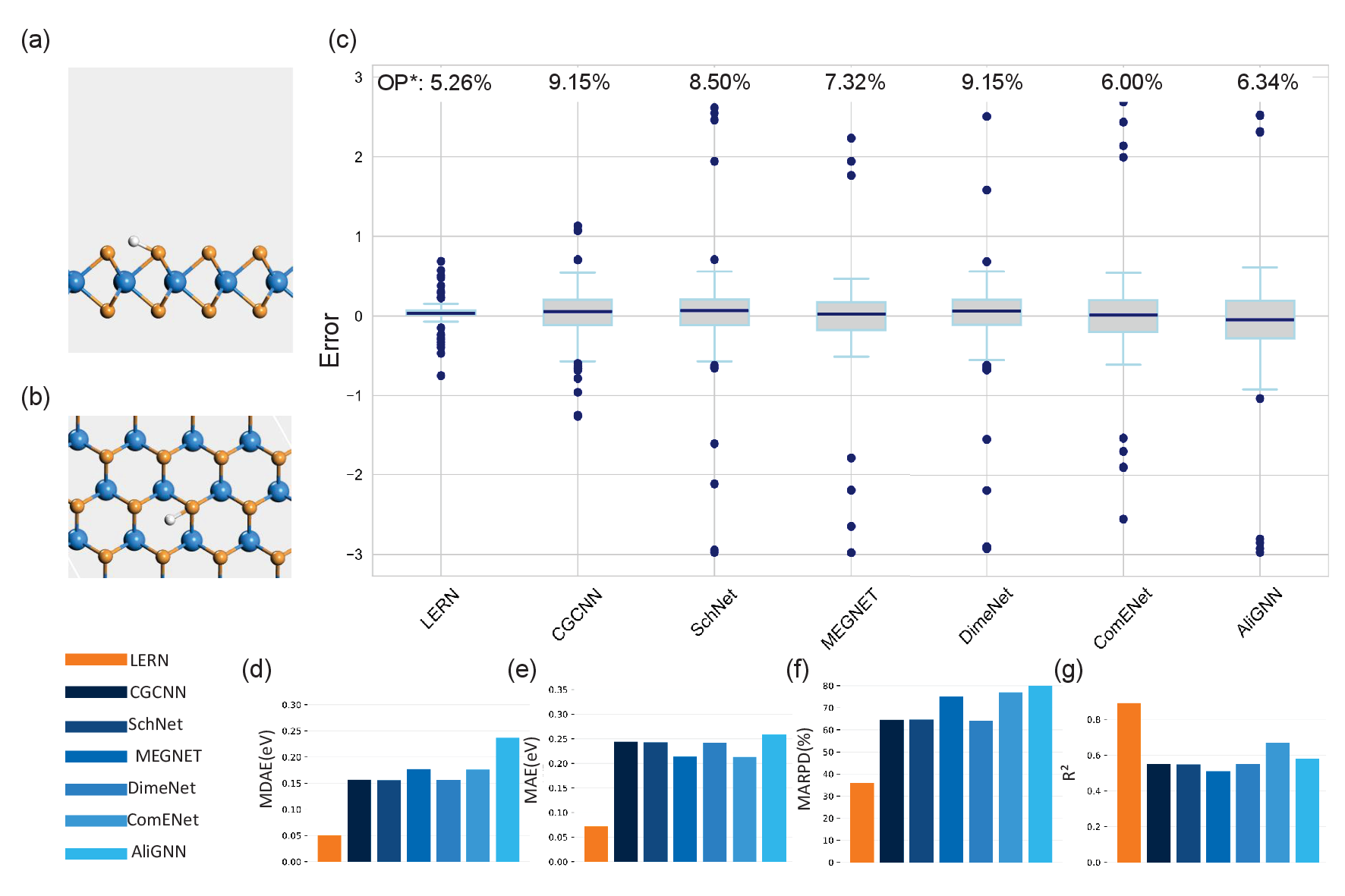}
\caption{(a)H adsorbed on 2D materials (b) and (c)Performance comparison of LERN with other representative models on 2D materials.(*op:Outlier percentage)}
\label{fig:box2}
\end{figure}

The LERN model still shows stable performance superiority on two-dimensional materials as showed in Figure \ref{fig:box2}. This is due to the fact that the feature engineering of the model only focuses on local information, independent of the material scale and thickness. The model also performs well on small datasets due to the introduction of a deep residual structure and the sharing of parameters. This also leads to the training time of LERN is faster than all other neural networks by an order of magnitude. It is noteworthy that all the aforementioned models were trained on a single RXT3080. When iterating 200 times, with the exception of LERN, the training time for all models exceeded two hours, while LERN required only approximately 26 minutes.  This is considerably faster than other conventional graph neural networks while the accuracy is greatly optimized.

\section{Discussion}

In this study, we have pioneered a cutting-edge local environment description method to refine the feature input of machine learning models. Leveraging an enhanced Voronoi Tessellation, we successfully extract geometric characteristics of 3D structures. This is further complemented by an innovative fingerprinting method that imparts localized information of adsorption sites to the model. This proposed descriptor has been applied to predict the adsorption energy of hydrogen atoms in catalytic processes, and our findings underscore that the local environment description method substantially outclasses prior models in prediction accuracy.

We have also broadened the scope of the proposed method by adapting it to deep learning algorithms, specifically by incorporating the MolCGCNN and LERN models. The experimental results affirm that our enhanced model offers superior prediction accuracy for smaller datasets and demands fewer computational resources. This underscores the efficacy of the local environment description approach. A notable innovation in our work is the employment of descriptors as graph inputs for the convolutional neural network, leading to a local environment residual network that markedly improves the final prediction accuracy by approximately 0.12 eV.

The marked advancement in prediction accuracy in our approach can be traced to the meticulous attention to and extraction of the local information of adsorption sites. By generating and predicting the adsorption energies of 1,283 adsorption sites for hydrogen atoms, based on the 2Dmatpedia database, we were able to identify 272 predicted adsorption energies within the range of plus and minus 0.5 eV after screening. Among these, seven have been previously reported in experiments, 69 in other computational studies, and 196 remain unprecedented in public literature or open databases. A significant highlight of our work is its superior speed compared to other models, made possible by extracting only the environment around the adsorbed atoms for training, thus notably enhancing screening efficiency. We are confident that this methodology can be expanded to a more diverse array of catalytic material discovery, fortifying the capability of machine learning models to predict the adsorption energy of various catalytic materials.

\section{Methods}

\subsection{Local environment fingerprint descriptors}

Using crystal graph to construct network frameworks has certain limitations. Firstly, the crystal graph structure only utilizes physical information such as atomic distances and angles, which makes it difficult to reflect the chemical properties between atoms. Additionally, the computational complexity of crystal graph networks is extremely high, with state-of-the-art models requiring single GPU computation times of up to hundreds of days. On the other hand, the arrangement and properties of atoms in the vicinity of the adsorption point are crucial factors influencing the energy of the hydrogen adsorption process\cite{strobel2006hydrogen}. Therefore, we propose to use faster machine learning algorithms to construct more effective descriptors for adsorption energy prediction, not only greatly increasing training speed but also achieving superior performance and accuracy. By simplifying the input and utilizing the powerful performance of neural networks, we can further improve the accuracy of adsorption energy prediction.

We have proposed a comprehensive framework for predicting adsorption energies, which involves several key steps, as shown in Fig.\ref{fig:ALL}. Firstly, we transform the original material structure into a Voronoi diagram, and extract improved local environment fingerprint information. Next, we further optimize this descriptor using machine learning techniques. Finally, we utilize the powerful performance of deep learning to construct, train, and predict using the model. This approach enables us to capture important physical features of the adsorption process and achieve high prediction accuracy.

We first improved the Crystallography Neural Network, but in practice we found that the construction of the Fingerprint draws on many previously successful methods for modelling chemical crystals. The atomic radius is a feature that better describes in vitro steric effects \cite{schutt2014represent}. However, atomic radii may also change due to changes in the environment. In the present work we use the atomic number instead of this feature as it is simple and deterministic.Pauling electronegativity has been shown to be a good feature of electron affinity\cite{li2017high}. To account for steric and ambient electron effects, the coordination number has been shown to be a successful feature\cite{calle2015introducing}. Crude estimates of the properties have proved successful and can improve predictive power, so we use the average adsorption energy as a description.

For the neighboring elements at each adsorption site, we utilized four elemental properties as follows: the atomic number of the element (Z), the Pauling electronegativity of the element ($\chi$), the number of neighboring atoms to which the element is coordinated to the adsorbate (CN), and the average adsorption energy of the element ($\Delta E$). The $\chi$ value was obtained from the Mendeleev database [29], and the $\Delta E$ value was calculated from the adsorption energy database, which represents the average of the adsorption energies of all catalysts containing this element. Additionally, we included the atom distance (D) to the adsorbate H, a parameter that is directly related to the adsorption energy magnitude in adsorption. Finally, we added the valence number (VN), which is calculated as the average of the elements within all the layers.

In addition, we note that the features generated by the general method all use the missing value filling method, where the individual vectors are the same for different elements and therefore do not reflect the information for each case. We therefore processed this by modifying the values of atomic number and Bowling electronegativity for each layer to the average of the layers. In this way, the fill values are different for each case. The relevant chemical properties of the cases can be better represented.

Thus, for the improved Fingerprint method, the two layers contain up to 7 elements, each corresponding to six parameters. Thus, a maximum of 42 feature inputs can be generated after modelling each case.
In addition, due to the presence of edge adsorption, we consider three layers of nearest neighbours, and experimentally, in surface adsorption, this method is able to adequately contain the information needed to calculate the adsorption energy. In addition, we use tools such as matminer to enrich the selection space of our features, which are finally filtered and validated using machine learning algorithms. The final result proves that we only need six descriptors to make the model reach the highest accuracy.

\subsection{Local environment descriptors for convolutional networks}

In lateset adsorption energy calculations, graph neural networks are typically utilized. Nevertheless, our local environment features can also be transformed into a 6*11 matrix, which meets all the necessary criteria for neural network utilization. As a result, we suggest a creative approach to feeding local environment descriptors into a convolutional neural network in matrix form. By doing so, we can leverage the advantages of deep learning in the context of adsorption energy calculations.

In this model, for each element E in a layer, there are six features f associated with it, which can be represented as:

\begin{equation}
E=[f_1,f_2,f_3,f_4,f_5,f_6]
\end{equation}

where each f corresponds to a descriptor mentioned earlier.

For a local environment with three layers and eleven elements, the input X can be represented as:

\begin{equation}
L=[E_i \vert i \in (1,11)]^T
\end{equation}

Here, L is a 6x11 matrix that serves as the input to a convolutional neural network, denoted as O(L), consisting of 32 residual blocks. Each block comprises two convolutional layers and a skip connection, which is used to construct the layers of ResNet. ResNet consists of multiple residual blocks, with four stages, each comprising 3, 4, 6, and 3 residual blocks, respectively, followed by a skip connection that adds the input X to the output of the second convolutional layer. The output of the n-th residual block can be written as:

\begin{equation}
Z_n = L + F_n(O_{n-1}(L))
\end{equation}

where $F_n$ represents the residual function of the n-th block, and $O_{n-1}(L)$ is the output of the (n-1)-th block. Finally, global average pooling and a fully connected layer are used to transform the feature map of the last layer into a scalar output:

\begin{equation}
O(L) = W_{out} Z_n + b_{out}
\end{equation}

where $W_{out}$ and $b_{out}$ are the weights and biases of the output layer, respectively.

The residual function $F_n$ can be defined as:

\begin{equation}
F_n(L) = \sigma(W_{2,n} \delta(W_{1,n} X + b_{1,n}) + b_{2,n})
\end{equation}

where $W_{1,n}$ and $W_{2,n}$ are the weights of two convolutional layers, $b_{1,n}$ and $b_{2,n}$ are their biases, $\delta$ represents the convolution operation, and $\sigma$ is the ReLU activation function.

Our framework possesses two key features to ensure the robustness and high performance of the model in adsorption tasks. Firstly, we adopt the mean absolute error as the loss function and train the network with the Adam optimization algorithm, ensuring the robustness and adaptability of the model. This choice not only adapts to adsorption tasks with different complex structures but also ensures that the model performs exceptionally well in various data scenarios. Secondly, we use the ResNet-34 model as our backbone network. ResNet-34 consists of 33 convolutional layers, providing depth and capability to the model, enabling it to be applied to larger databases, thereby enhancing the model's accuracy and robustness. ResNet-34 is a widely used deep learning framework, whose outstanding performance has been proven in numerous fields. Our choice also provides robust support for adsorption tasks. Overall, by constructing a local environment convolutional network, we integrate the high-importance features of the three nearest neighbor atoms surrounding the adsorbate and learn the relationships between each layer's elements. Then, we learn from a large dataset's data based on the convolutional network. Introducing residual networks reduces the model's learning difficulty and enhances its generalizability.

\subsection{Regression result evaluation}

To train the model, several regression outcome evaluation methods are employed. The true and predicted values in the experiment are represented as follows:

True value: \begin{equation}
\hat{y}=\left\{ \hat{y}_1\, {\hat{y}}_2\cdots, \hat{y}_n \right\}\label{eq1}
\end{equation}
Predicted value:
\begin{equation}
y=\left\{y_1,y_2,\cdots,y_n\right\}\label{eq2}
\end{equation}

Performance metrics for all methods used in this article, which include: Median Absolute Error (MDAE), Mean Absolute Error (MAE), Root Mean Squared Error (RMSE), Mean Absolute Relative Percent Difference (MARPD).

The median absolute error is particularly interesting because it is robust to
outliers, unit of this indicator is eV . The median absolute error estimated over n samples is defined as follows:
\begin{equation}
{\rm MDAE}(y,\hat{y})=median(\vert y_1-\hat{y}_1\vert,\cdots \vert y_n-\hat{y}_n \vert )
\end{equation}

Mean Absolute Relative Percent Difference (MARPD) is used because it provides
normalized measures of accuracy that may be more interpretable for those unfamiliar
with adsorption energy measurements in eV .
\begin{equation}
{\rm MARPD} = \frac { 1 } { n } \sum _ { i = 1 } ^ { n } \left\vert \frac { y _ { i }-\hat y _i} { \vert y _ { i  } \vert + \vert \hat y _i \vert}\cdot 100\%  \right\vert
\end{equation}

\newpage
\citep{Bibliography}

\end{document}